\documentclass[runningheads]{llncs}

\usepackage[strings]{underscore}
\usepackage{graphicx}
\usepackage[normalem]{ulem}
\usepackage{adjustbox}
\usepackage{hhline}
\usepackage{multirow}
\usepackage{url}
\usepackage{rotating}
\usepackage[cmex10]{amsmath}
\usepackage{wrapfig}
\usepackage{tabularx}
\usepackage{array}

\newcolumntype{Y}{>{\centering\arraybackslash}X}

\newcommand{\RQonetwo}{Are traditional mutation operators enough to prevent null-type faults?}
\newcommand{\RQthree}{To what extent is the addition of null-type mutation operators useful in practice?}

\begin{document}
\mainmatter
\title{Do Null-Type Mutation Operators\\ Help Prevent Null-Type Faults?}
\titlerunning{Do Null-Type Mutation Operators Help Prevent Null-Type Faults?}
			
\author{Ali Parsai\inst{1}\orcidID{0000-0001-8525-8198} \and
        Serge Demeyer\inst{2}\orcidID{0000-0002-4463-2945}}
            
\authorrunning{Ali Parsai\and Serge Demeyer}
			
\institute{University of Antwerp\\
\email{ali.parsai@uantwerpen.be}
\and
University of Antwerp and Flanders Make\\
\email{serge.demeyer@uantwerpen.be}
}

\maketitle

\begin{abstract}
The null-type is a major source of faults in Java programs, and its overuse has a severe impact on software maintenance. Unfortunately traditional mutation testing operators do not cover null-type faults by default, hence cannot be used as a preventive measure. We address this problem by designing four new mutation operators which model  null-type faults explicitly. We show how these mutation operators are capable of revealing the missing tests, and we demonstrate that these mutation operators are useful in practice. For the latter, we analyze the test suites of 15 open-source projects to describe the trade-offs related to the adoption of these operators to strengthen the test suite. 

\keywords{	Software Maintenance,
	Software Testing,
	Mutation Testing,
	Null-Type,
	Test Quality
}
\end{abstract}

\section{Introduction}
\label{S:Intro}

The null-type is a special type in Java that has no name, cannot be casted, and practically equates to a literal that can be of any reference type~\cite{Gosling2014}. The null-type is commonly misused, and frequently reported and discussed as an issue by developers~\cite{Osman2016}. The null-type is the source of the majority of faults in Java programs~\cite{Osman2014}, and its overuse has a severe impact on software maintenance~\cite{Kimura2014}. 
On the one hand, this scenario should push developers to build test suites capable of identifying null-type faults. On the other hand, developers without specific test requirements may struggle to identify all  code elements or properties that the test must satisfy. To address this problem, we propose mutation testing as a way for improving the test suite to handle potential null-type faults.

Mutation testing is a technique to measure the quality of a test suite by assessing its fault detection capabilities~\cite{DeMillo1978}. 
Mutation testing is a two-step process. First, a small syntactic change is introduced in the production code. This change is obtained by applying a ``mutation operator'', and the resulting changed code is called a ``mutant''. Then, the test suite is executed for that mutant; if any of the tests fail, the mutant is ``killed'', otherwise, the mutant has ``survived''. 
Herein lies the aspect of mutation testing that we want to exploit: the identification of survived mutants that need to be killed. %

Mutation operators are modeled after the common developer mistakes~\cite{Just2014}. Over the years,  multiple sets of mutation 
operators have been created to fit in different domains. By far the most commonly used mutation operators are the ones 
introduced in Mothra by Offutt et al.~\cite{Offutt1996}. They use 10 programs written in Fortran to demonstrate that their 
reduced-set mutation operators is enough to produce a mutation-adequate test suite  that can kill almost all of the mutants 
generated by the mutation operators of the complete-set.  Later on, several attempts have been made to extend Offutt's mutation 
operators, for instance, to cope with the specificities of object-oriented programming~\cite{Ma2002}. Yet, none of the proposed 
mutation operators explicitly model null-type faults. As a result, mature general-purpose mutation testing tools currently used in literature, such as PITest~\cite{Coles2016} and Javalanche~\cite{Schuler2009a}, do not cope explicitly with this type of faults by default. Therefore, the created mutants \textit{risk} not  being adequate to derive test requirements that handle null-type faults. Whether this risk is concrete or not depends on the ability of the available mutation operators to account for these faults. Yet, no study has explored this aspect.

This paper investigates the usefulness of mutation operators able to model null-type faults in 
order to strengthen the test suite against these faults. %
For this reason, we introduce four new mutation operators related to null-type faults. These mutation operators are modeled to cover the typical null-type faults introduced by developers~\cite{Osman2016}. We incorporate these mutation operators in LittleDarwin, an extensible open-source tool for mutation testing~\cite{Parsai2017}, creating a new version called  LittleDarwin-Null. 
We organize our research in two steps: we show that (i) the current general-purpose mutation testing tools do not account for null-type faults by default, and modeling operators for null-type faults can drive the improvement of the test suite in practice, and (ii) the test suites of real open-source projects cannot properly catch null-type faults.
The paper is driven by the following research questions:
\begin{itemize}
	\item \textbf{RQ1:} \textit{\RQonetwo}

\item \textbf{RQ2:} \textit{\RQthree}

\end{itemize}

The rest of the paper is organized as follows: 
In Section~\ref{S:Background}, background information and  related work is provided. 
In Section~\ref{S:CSDesign}, the details of the experiment are discussed.
In Section~\ref{S:Results}, the results are analyzed. 
In Section~\ref{S:Threats}, we discuss the threats that affect the results. 
Finally, we present the  conclusion in Section~\ref{S:Conclusion}.

              \vspace{-1em}
\section{Background and Related Work}
\label{S:Background}

Mutation testing is the process of injecting faults into a software system and then verifying whether the test suite indeed fails, and thus detects the injected fault. %
First, a faulty version of the software is created by introducing faults into the system \textit{(Mutation)}. This is done by applying a transformation \textit{(Mutation Operator)} on a certain part of the code.  After generating the faulty version of the software \textit{(Mutant)}, it is passed onto the test suite. If a test fails, the mutant is marked as killed \textit{(Killed Mutant)}. If all tests pass, the mutant is marked as survived \textit{(Survived Mutant)}.

\textbf{Mutation Operators.}
A mutation operator is a transformation which introduces a single syntactic change into its input. The first set of mutation operators were reported in King et al.~\cite{King1991}. These mutation operators work on essential syntactic entities of programming languages such as arithmetic, logical, and relational operators.
For object-oriented languages, new mutation operators were proposed~\cite{Ma2002}.
The mature mutation testing tools of today  still mostly use the \textit{traditional} (i.e. method-level) mutation operators~\cite{Papadakis2018}.

\textbf{Equivalent Mutants.}
An \emph{equivalent mutant} is a mutant that does not change the semantics of the program, i.e. its output is the same as the original program for any possible input. Therefore, no test case can differentiate between an equivalent mutant and the original program. The detection of equivalent mutants is undecidable due to the halting problem~\cite{Offutt1997}. %

\textbf{Mutation Coverage.} 
Mutation testing allows software engineers to monitor the fault detection capability of a test suite by means of mutation coverage~\cite{Jia2011}. 
A test suite is said to achieve \textit{full mutation test adequacy} whenever it can kill all the non-equivalent mutants, thus reaching a mutation coverage of 100\%. Such test suite is called a \textit{mutation-adequate test suite}.

\textbf{Mutant Subsumption.}
Mutant subsumption is defined as the relationship between two mutants \texttt{A} and \texttt{B} in which \texttt{A} subsumes \texttt{B} if and only if the set of inputs that kill \texttt{A} is guaranteed to kill 
\texttt{B}~\cite{Kurtz2015}. The subsumption relationship for faults has been defined by Kuhn in 1999~\cite{Kuhn1999}. Later 
on, Ammann et al. tackled the theoretical side of mutant subsumption~\cite{Ammann2014} where they define \textit{dynamic} mutant subsumption as follows: Mutant \texttt{A} dynamically subsumes Mutant \texttt{B} if and only if (i) \texttt{A} is killed, and (ii) every test that kills \texttt{A} also kills \texttt{B}.
The main purpose behind the use of mutant subsumption is to  detect redundant mutants. These mutants create multiple threats to the validity of mutation analysis~\cite{Papadakis2016}. This is done by determining the dynamic subsumption relationship among a set of mutants, and keep only those that are not subsumed by any other mutant. 

\textbf{Mutation Testing Tools.}
In this study, we use three different mutation testing tools: Javalanche, PITest, and LittleDarwin.
Javalanche is a mutation testing framework for Java programs that attempts to be efficient, and not produce equivalent mutants~\cite{Schuler2009a}. It uses byte code manipulation in order to speed up the process of mutation testing. 
Javalanche has been used in numerous studies in the past (e.g. \cite{Gligoric2013,Fraser2012}).
PITest is a state-of-the-art mutation testing system for Java, designed to be fast and scalable~\cite{Coles2016}. 
PITest is the de facto standard for mutation testing within Java, and it is used as a baseline in mutation testing research (e.g.~\cite{Inozemtseva2014,Parsai2014}).
LittleDarwin is a mutation testing tool designed to work out of the box with complicated industrial build systems. For this, it has a loose coupling with the test infrastructure, instead relying on the build system to run the test suite. %
LittleDarwin has been used in several studies, and is capable of performing mutation testing on complicated software systems~\cite{Parsai2015,Parsai2016,Parsai2016M}. %
For more information about LittleDarwin please refer to Parsai et al.~\cite{Parsai2017}.
We implemented the new null-type mutation operators in a special version of LittleDarwin called LittleDarwin-Null. LittleDarwin and LittleDarwin-Null only differ in the set of mutation operators used, and are identical otherwise.

\textbf{Related Work.}
Creating new mutation operators to deal with the evolution of software languages is a trend in mutation testing research. For example, mutation operators have been designed to account for concurrent code~\cite{Bradbury2006}, aspect-oriented programming~\cite{Ferrari2008}, graphical user interfaces~\cite{Oliveira2015}, modern C++ constructs~\cite{Parsai2018}, and Android applications~\cite{Deng2017}. Nanavati et al. have previously studied mutation operators targeting memory-related faults~\cite{Nanavati2015}. However, the difference in the semantics of null object of Java and NULL macro of C is sufficient to grant the need for a separate investigation.  

\vspace{-1em}
\section{Experimental Setup}
\label{S:CSDesign}
In this section,  we first introduce our proposed mutation operators, and then we discuss the  experimental setup we used to address our research questions.

              \vspace{-1em}
\subsection{Null-Type Mutation Operators}
\label{SS:Mutators}
We derived four null-type mutation operators to model the typical null-type faults often encountered by developers~\cite{Osman2014}. 
These mutation operators are presented in Table~\ref{table:nulloperators}.

\begin{table}
\vspace{-5mm}
\centering
\caption{Null-Type Faults and Their Corresponding Mutation Operators}    
\label{table:nulloperators}
    \adjustbox{max width=\linewidth}{

\begin{tabular}{|l|l|}
	
	\hline \textbf{Mutation Operator}  & \textbf{Description}  \\ 
	\hline \hline  NullifyReturnValue & If a method returns an object, it is replaced by \texttt{null} \\ 
	\hline NullifyInputVariable & If a method receives an object reference, it is replaced by \texttt{null} \\ 
	\hline  NullifyObjectInitialization & Wherever there is a \texttt{new} statement, it is replaced with \texttt{null} \\ 
	\hline  NegateNullCheck  & Any binary relational statement containing \texttt{null} at one side is negated  \\
	\hline 
\end{tabular}

 }

\end{table}

              \vspace{-3em}
\subsection{Case Study}
For RQ1, we use a didactic project. 
For RQ2, %
we use 15 open-source projects.

\textbf{RQ1.}
In order to address RQ1, we chose a modified version of  VideoStore  as a small experimental 
project~\cite{Fowler1999}. 
Choosing a small project allows us to (i) create a mutation-adequate test suite ourselves, (ii) find out which mutants are equivalent, and (iii) avoid complexities when using multiple mutation testing tools. The source code for VideoStore is available in the replication package.

\begin{table}
	\centering
	\caption[]{Projects Sorted by Mutation Coverage}
	\label{table:cases}
	
	\begin{tabular}{|c|c|c|c|c|c|c|c|c|}
		\hline \multirow{2}{*}{\textbf{Project}} & \multirow{2}{*}{\textbf{Ver.}} & \multicolumn{2}{c|}{\textbf{Size (LoC)}} & \multirow{2}{*}{\textbf{\#C}} & \multirow{2}{*}{\textbf{TS}} & \multirow{2}{*}{\textbf{SC}} & \multirow{2}{*}{\textbf{BC}} & \multirow{2}{*}{\textbf{MC}}  \\
		\hhline{~~--~~~~} &  & \textbf{Prod.} & \textbf{Test} &  &  & & &  \\ 
		\hline
		\hline Apache Commons CLI & 1.3.1   & 2,665 & 3,768  & 816 & 15 & 96\% & 93\% & 94\% \\ 
		\hline JSQLParser & 0.9.4 & 7,342 & 5,909  & 576 & 19 & 81\% & 73\% & 94\% \\ 
		\hline jOpt Simple & 4.8 &  1,982 & 6,084 & 297 & 14 & 99\% & 97\% & 92\%\\ 
		\hline Apache Commons Lang & 3.4   & 24,289 & 41,758 & 4,398 & 30 & 94\% & 90\% & 91\%  \\ 
		\hline Joda Time & 2.8.1  & 28,479 & 54,645 & 1,909 & 42 & 90\% & 81\% & 82\% \\ 
		\hline Apache Commons Codec & 1.10   & 6,485 & 10,782 & 1,461 & 10  & 96\% & 92\% & 82\% \\ 
		\hline Apache Commons Collections& 4.1 & 27,914 & 32,932  & 2,882 & 26 &  85\% & 78\%  & 81\%\\
		\hline VRaptor& 3.5.5 & 14,111 & 15,496  & 3,417 & 65 & 87\% & 81\% & 81\% \\ 
		\hline HTTP Request & 6.0 & 1,391 & 2,721   & 446 & 15 & 94\% & 75\% & 78\% \\ 
		\hline Apache Commons FileUpload & 1.3.1 & 2,408 & 1,892   & 846 & 19 & 76\% & 74\% & 77\% \\ 
		\hline jsoup & 1.8.3 & 10,295 & 4,538  & 888 & 43 & 82\% & 72\% & 76\% \\ 
		\hline JGraphT & 0.9.1 & 13,822 & 8,180  & 1,150& 31 & 79\% & 73\% & 69\% \\ 
		\hline PITest & 1.1.7 & 17,244 & 19,005  & 1,044 & 19 &  79\% & 73\%  & 63\% \\ 
		\hline JFreeChart & 1.0.17 & 95,354 & 41,238  & 3,394 & 4 &  53\% & 45\%  & 35\%\\ 
		\hline PMD & r7706 & 70,767 & 43,449 & 7,706 & 20 & 62\% & 54\% & 34\% \\  
		\hline
					\multicolumn{9}{r}{\scriptsize Acronyms: Version (Ver.), Line of code (LoC), Production code (Prod.), Number of commits (\#C),} \\
					\multicolumn{9}{r}{\scriptsize Team size (TS), Statement coverage (SC), Branch coverage (BC), Mutation coverage (MC)}\\
	\end{tabular} 
	              \vspace{-1em}
	
\end{table}

\textbf{RQ2.}
We selected 15 open-source projects for our empirical study (Table~\ref{table:cases}). 
The selected projects differ in size of their production code and  test code, number of commits, and team size  to provide a wide range of possible scenarios. Moreover, they also differ in the adequacy of their test suite based on statement, branch, and mutation coverage (Table~\ref{table:cases}). We used JaCoCo and Clover %
for statement and branch coverage, and LittleDarwin for mutation coverage.

 \vspace{-1em}
\section{Results and Discussion}
\label{S:Results}

\vspace{-1em}
\textbf{\small RQ1: \RQonetwo}
 
We are interested to compute the number of killed, survived and equivalent mutants along with three versions of VideoStore. The first version we analyze is the original one (VideoStore Orig). This version has only 4 tests.  
Then, we create a mutation-adequate test suite that kills all mutants generated by the general-purpose tools (Javalanche, PITest, and LittleDarwin).  In
this version (VideoStore TAdq) we added 15 tests. 
Finally, we create a mutation-adequate test suite that kills all mutants, included the ones generate by LittleDarwin-Null. In this version (VideoStore NAdq) we added 3 more tests.

 \begin{table}
     \centering
     \caption{Mutation testing results for VideoStore}
     \label{table:RQ1}
     
     \adjustbox{max width=\linewidth}{  	\begin{tabularx}{\textwidth}{YYYYYYYYYYYYY}
             \hline
             \multicolumn{1}{|Y|}{\multirow{2}{*}{\textbf{Program}}} & \multicolumn{3}{c|}{\textbf{LittleDarwin}}   & \multicolumn{3}{c|}{\textbf{PITest}}      & \multicolumn{3}{c|}{\textbf{Javalanche}}    & \multicolumn{3}{Y|}{\textbf{LittleDarwin-Null}}  \\ \cline{2-13} 
             \multicolumn{1}{|c|}{}                         & \multicolumn{1}{Y|}{\textbf{K}}  & \multicolumn{1}{Y|}{\textbf{S}}  & \multicolumn{1}{Y|}{\textbf{E}} & \multicolumn{1}{Y|}{\textbf{K}}  & \multicolumn{1}{Y|}{\textbf{S}}  & \multicolumn{1}{Y|}{\textbf{E}} & \multicolumn{1}{Y|}{\textbf{K}}   & \multicolumn{1}{Y|}{\textbf{S}}  & \multicolumn{1}{Y|}{\textbf{E}} & \multicolumn{1}{Y|}{\textbf{K}}  & \multicolumn{1}{Y|}{\textbf{S}}  & \multicolumn{1}{Y|}{\textbf{E}} \\ \hline \hline
             \multicolumn{1}{|c|}{VideoStore Orig}          & \multicolumn{1}{c|}{24} & \multicolumn{1}{c|}{18} & \multicolumn{1}{c|}{2} & \multicolumn{1}{c|}{25} & \multicolumn{1}{c|}{43} & \multicolumn{1}{c|}{5} & \multicolumn{1}{c|}{87}  & \multicolumn{1}{c|}{69} & \multicolumn{1}{c|}{11} & \multicolumn{1}{c|}{11} & \multicolumn{1}{c|}{14} & \multicolumn{1}{c|}{1} \\ \hline
             \multicolumn{1}{|c|}{VideoStore TAdq}          & \multicolumn{1}{c|}{42} & \multicolumn{1}{c|}{0}  & \multicolumn{1}{c|}{2} & \multicolumn{1}{c|}{68} & \multicolumn{1}{c|}{0}  & \multicolumn{1}{c|}{5} & \multicolumn{1}{c|}{202} & \multicolumn{1}{c|}{0}  & \multicolumn{1}{c|}{11} & \multicolumn{1}{c|}{22} & \multicolumn{1}{c|}{3}  & \multicolumn{1}{c|}{1} \\ \hline
             \multicolumn{1}{|c|}{VideoStore NAdq}          & \multicolumn{1}{c|}{42} & \multicolumn{1}{c|}{0}  & \multicolumn{1}{c|}{2} & \multicolumn{1}{c|}{68} & \multicolumn{1}{c|}{0}  & \multicolumn{1}{c|}{5} & \multicolumn{1}{c|}{202} & \multicolumn{1}{c|}{0}  & \multicolumn{1}{c|}{11} & \multicolumn{1}{c|}{25} & \multicolumn{1}{c|}{0}  & \multicolumn{1}{c|}{1} \\ \hline
             \multicolumn{13}{r}{K: Killed, S: Survived, E: Equivalent}                                                                                                                                                                                                                                                                                                         
            \end{tabularx}}
                          \vspace{-2em}
        \end{table}
 
Table~\ref{table:RQ1} shows the number of remaining mutants after each phase of test development: 
VideoStore Orig, VideoStore TAdq, and   VideoStore NAdq. The discrepancy in total number of generated mutants for the three versions of the program in case of Javalanche is due to its particular optimizations.
In VideoStore Orig, there are several survived mutants according to all the tools.  This is because the test suite accompanying the VideoStore program was not adequate.

\begin{wrapfigure}[19]{r}{0.62\linewidth}
    \centering
    \fbox{\includegraphics[width=1\linewidth]{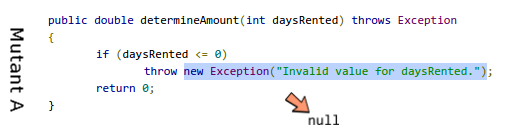}}
    \fbox{\includegraphics[width=1\linewidth]{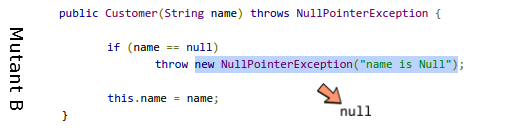}}
    \fbox{\includegraphics[width=1\linewidth]{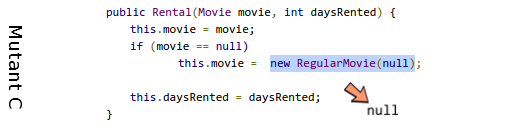}}
    \caption{The Surviving Non-Equivalent Null-Type Mutants}
    \label{fig:nullmutants}
\end{wrapfigure}

\begin{figure}
	\centering
	\fbox{\includegraphics[width=0.7\linewidth]{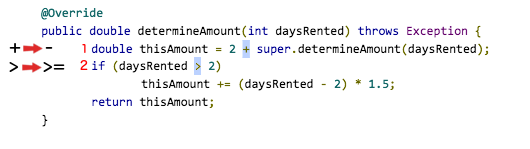}}
	\caption{Two of the Equivalent Mutants Generated by Traditional Mutation Operators}
	\label{fig:pitestequivalentexample}
	\vspace{-2em}
\end{figure}

In VideoStore TAdq, we create a mutation-adequate  version of the test suite  with respect to the results of PITest, Javalanche, and LittleDarwin. In the process of creating this test suite, we noticed that all of these tools produce equivalent mutants. Two of such mutants are shown in Figure~\ref{fig:pitestequivalentexample}. Mutant~A is equivalent because the method  \texttt{super.determineAmount} always returns 0, so it does not matter whether it is added to or subtracted from \texttt{thisAmount}. Mutant~B is also equivalent, because if \texttt{daysRented} is 2, the value added to \texttt{thisAmount} is 0. %
We analyzed VideoStore TAdq with LittleDarwin-Null in order to find out whether the mutation-adequate test suite according to three general-purpose tools is able to kill all the null-type mutants. 
By analyzing the 26 generated mutants, we noticed that 22 mutants were killed and 4 survived. The manual review of these mutants show that one of them is an equivalent mutant. 

	\vspace{-2em}

              \vspace{-2em}
\begin{figure}
	\centering
	\fbox{\includegraphics[width=0.7\linewidth]{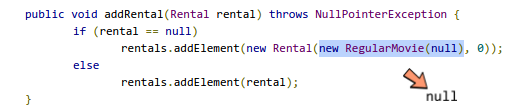}}
	\caption{One of the Equivalent Mutants Generated by Null-Type Mutation Operators}
	\label{fig:nullequivalentexample}
	\vspace{-2em}
\end{figure}

Considering that 3 mutants generated by null-type mutation operators are not equivalent, and yet the mutation-adequate test suite we created according to the general-purpose tools cannot kill them, we conclude 
that \textbf{using traditional mutation operators to strengthen the test suite does not necessarily prevent null-type faults}.

The four mutants survived in VideoStore TAdq are all of type \textit{NullifyObjectInitialization}. 
Figure~\ref{fig:nullequivalentexample} shows the equivalent null-type mutant. Here the default behavior of \texttt{Rental} object is to create a new \texttt{RegularMovie} object when it receives \texttt{null} as its input. So, replacing \texttt{new~RegularMovie(null)} with null does not change the behavior of the program. %

              \vspace{-0.1em}

\begin{wrapfigure}[28]{r}{0.65\linewidth}
	\vspace*{-2em}
	\centering
	\fbox{\includegraphics[width=1\linewidth]{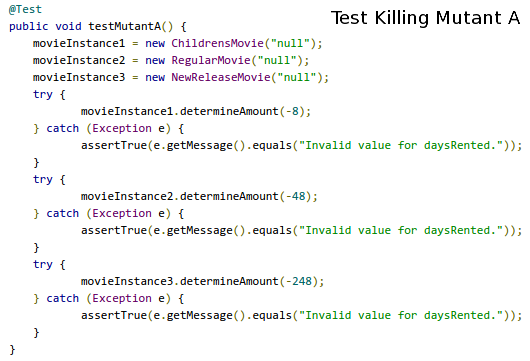}}
	\fbox{\includegraphics[width=1\linewidth]{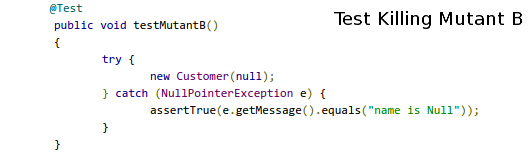}}
	\fbox{\includegraphics[width=1\linewidth]{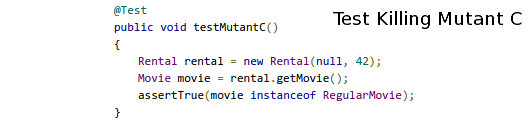}}
	\caption{The Tests Written to Kill the Surviving Null-Type Mutants}
	\label{fig:nullmutantstests}
\end{wrapfigure}

The three remaining surviving mutants are described in Figure~\ref{fig:nullmutants}. Here, 
mutants A and B replace the exception with \texttt{null}. Consequently, as opposed to the program throwing a detailed exception, the mutant always throws an empty \texttt{NullPointerException}. \\
Such a mutant is desirable to kill, since %
 the program would be able to throw an unexpected exception due to a fault that the test suite cannot recognize. 
In the case of Mutant C, it replaces the initialization of a \texttt{RegularMovie} object with \texttt{null}. This means that as opposed to the program that guarantees the private attribute \texttt{movie} is always instantiated, 
the same attribute contains a null literal  in the mutant. If not detected, a \texttt{NullPointerException} might be thrown 
when another object tries to access the \texttt{movie} attribute of this object. 

We created three new tests to kill each of the survived mutants. These tests are shown in Figure~\ref{fig:nullmutantstests}. Here,  \texttt{testMutantA} and \texttt{testMutantB} verify whether the unit under test throws  the correct exception if called with an invalid input value. \texttt{testMutantC} verifies whether the unit under test is able to handle a null input correctly. These three tests are not ``happy path tests'', namely a well-defined test case using known input, which executes without exception and produces an expected output. Consequently, they might not be intuitive for a test developer to consider, even though they are known as good testing practice~\cite{Alexander2003}. 
If not for the three survived null-type mutants, these tests would not have been written. This leads us to conclude that \textbf{traditional mutation operators are not enough to prevent null-type faults.}

 \textbf{\small RQ2: \RQthree}
 
 RQ1 shows for the VideoStore project that mutation testing tools need to introduce explicit mutation operators for modeling null-type faults.
 Yet, such a project  is not representative of real projects. 
 In this RQ, we want to verify to what extent null-type mutation operators are useful \textit{in practice}. 
 For this reason, we perform an experiment that involves real open-source projects. 
 After introducing null-type mutants, two groups of mutants are affected:  
 (i) survived mutants are the targets the developer needs during test development, 
 (ii) killed mutants show the types of faults the test suite can already catch.

Considering this,
we can justify the effort needed for extending mutation testing by incorporating null-type mutants only if:
(i) the real test suites do not already kill most of the null-type mutants, 
(ii) the null-type mutants are not increasing redundancy by a large margin.
Otherwise, the current mutation testing tools %
are already ``good enough" for preventing null-type faults.

 To verify to what extent the null-type mutants ``do matter"  when testing for null-type faults we analyze both killed and survived mutants: %

In case of survived mutants,  we analyze the number of survived mutants that each mutation operator generates for each project. We divide this analysis into two parts. 
First, we analyze survived mutants for null-type and traditional mutation operators.  
Second, we analyze each mutation operator individually to find out which one produces the most surviving mutants. 
This analysis shows whether the survived mutants produced by the null-type mutation operators are ``enough'' to  drive the test development process.  
 
In case of killed mutants, we take all projects as a whole, and we analyze whether the killed null-type mutants are redundant when used together with traditional mutation operators.
We measure redundancy using   %
dynamic mutant subsumption: we analyze the distributions of  subsuming, killed, and all null-type mutants. 
This way we can tell whether or not the null-type mutation operators are producing ``valuable'' mutants to strengthen the test suite. %

    \begin{table}[t]
        \centering 
        \caption{Mutants Generated by LittleDarwin and LittleDarwin-Null}
        \label{table:RQ3} 
        \adjustbox{max width=\linewidth}{ \begin{tabular}{|c|c|c|c|c|c|c|}
                \hline 
                \multirow{2}{*}{\textbf{Project}} & \multicolumn{3}{c|}{\textbf{Traditional Mutation Operators}} & \multicolumn{3}{c|}{\textbf{Null-Type Mutation Operators}} \\ 
                \cline{2-7} & \textbf{Survived} & \textbf{Killed} & \textbf{Total} & \textbf{Survived} & \textbf{Killed} & \textbf{Total} \\ 
                \hline \hline Apache Commons CLI & 24 & 318 & 342 & 71 & 415 & 486 \\ 
                \hline JSQLParser & 31 & 457 & 488 & 358 & 1,062 & 1,420 \\ 
                \hline jOpt Simple & 17 & 189 & 206 & 37 & 494 & 531 \\ 
                \hline Apache Commons Lang & 559 & 5,455 & 6,014 & 564 & 5,469 & 6,033 \\ 
                \hline Joda Time & 892 & 3,978 & 4,870 & 836 & 5,371 & 6,207 \\ 
                \hline Apache Commons Codec & 364 & 1,612 & 1,976 & 147 & 927 & 1,074 \\ 
                \hline Apache Commons Collections & 638 & 2,705 & 3,343 & 1,179 & 5,851 & 7,030 \\ 
                \hline VRaptor & 111 & 478 & 589 & 795 & 2,111 & 2,906 \\ 
                \hline HTTP Request & 49 & 178 & 227 & 69 & 383 & 452 \\ 
                \hline Apache Commons FileUpload & 81 & 273 & 354 & 137 & 211 & 348 \\ 
                \hline jsoup & 291 & 928 & 1,219 & 553 & 1,455 & 2,008 \\ 
                \hline JGraphT & 416 & 940 & 1,356 & 834 & 1,457 & 2,291 \\ 
                \hline PITest & 398 & 672 & 1,070 & 551 & 2,964 & 3,515 \\ 
                \hline JFreeChart & 10,558 & 5,603 & 16,161 & 8,563 & 6,248 & 14,811 \\ 
                \hline PMD & 5,205 & 2,734 & 7,939 & 5,099 & 4,613 & 9,712 \\ 
                \hline 
                \hline \textbf{Total} &  \textbf{19,634} & \textbf{26,520} & \textbf{46,154} & \textbf{19,793} & \textbf{39,031} & \textbf{58,824} \\
                \hline
                
            \end{tabular} }
                           \vspace{-2em}
        \end{table}

\textit{Survived mutants.} Table~\ref{table:RQ3} shows for each project the number of survived, killed, and total generated mutants for both groups of mutation operators. %
The first noticeable trend is a strong correlation ($ R^2 = 0.81 $) between survived to killed ratio (SKR) of the traditional mutants and SKR of the null-type mutants.
One exception to this trend is JSQLParser, in which there are significantly more survived null-type mutants than  survived traditional mutants. Investigating further, we find  that this happens because 50 small classes lack statements that can be mutated by the traditional mutation operators. However,  null-type mutation operators are able to generate mutants for these classes. This uncovers many of the weaknesses of the test suite.
On the other side of the fence, there is PITest, in which a single class (\texttt{sun.pitest.CodeCoverageStore}) contains  many arithmetic operations  while poorly tested, so it produces 129 out of 398  survived traditional mutants. This shows that \textbf{the usefulness of the null-type mutation operators is program-dependent.}   
            
                 \begin{figure}
                     \centering
                     \includegraphics[width=0.90\linewidth]{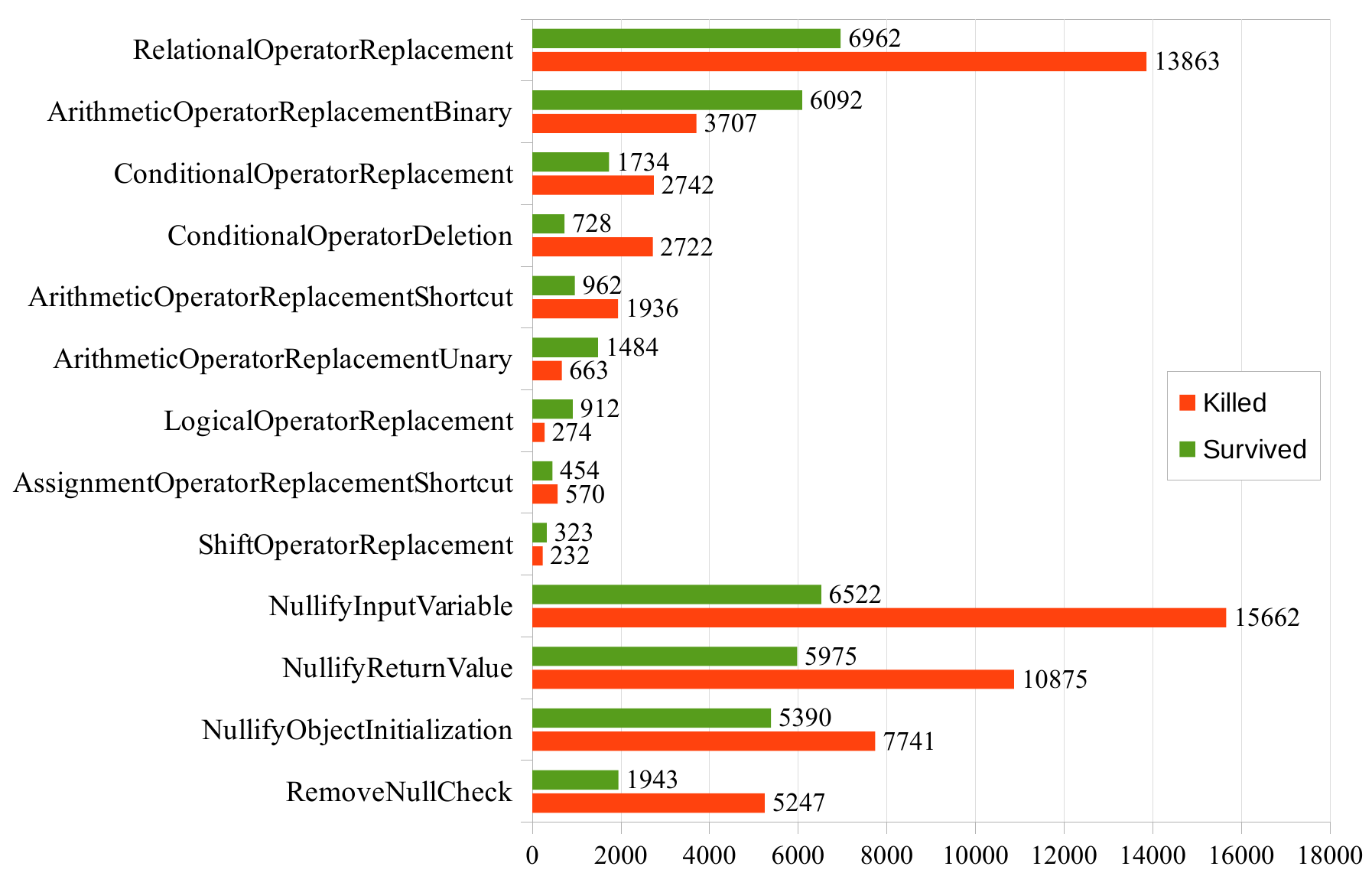}
                     \caption{Number of killed and survived mutants for  each mutation operator}
                     \label{fig:numberofmutantsperoperator}
                     \vspace{-1em}
                    \end{figure}

Figure~\ref{fig:numberofmutantsperoperator} shows the number of killed and survived mutants for each mutation operator. We see that among the traditional mutation operators,  \textit{ArithmeticOperatorReplacementBinary}, \textit{LogicalOperatorReplacement}, and \textit{ArithmeticOperatorReplacementUnary} have the highest ratio of survived to killed mutants. This means that these mutation operators are generating mutants that are harder to kill than the rest. The same can be observed among the null-type mutation operators, where \textit{NullifyObjectInitialization} produces harder to kill mutants than the others. %
This is as we expected, since \textit{NullifyInputVariable} applies a major change to the method (removal of an input), and \textit{NegateNullCheck} negates a check that the developer deemed necessary. However, the unexpected part of the result is that so many of the mutants generated by \textit{NullifyReturnValue} have survived. This means that lots of methods are not tested on their output correctly. This can be due to the fact that many of such methods are not tested directly, and when tested indirectly, their results only affect a small part of the program state of the method under test.

In general, the number of survived null-type mutants has a strong correlation with the number of survived traditional mutants for most projects. This implies that not all parts of the code are tested well. However, the exceptions to this rule are caused by classes that produce many more mutants of a particular type. Here, our results show that \textbf{the null-type mutation operators complement the traditional mutation operators and vice versa by each providing a large portion of survived mutants}.

\begin{wrapfigure}[12]{r}{0.50\linewidth}
                \vspace{-2em}
                \centering
                \includegraphics[width=1\linewidth]{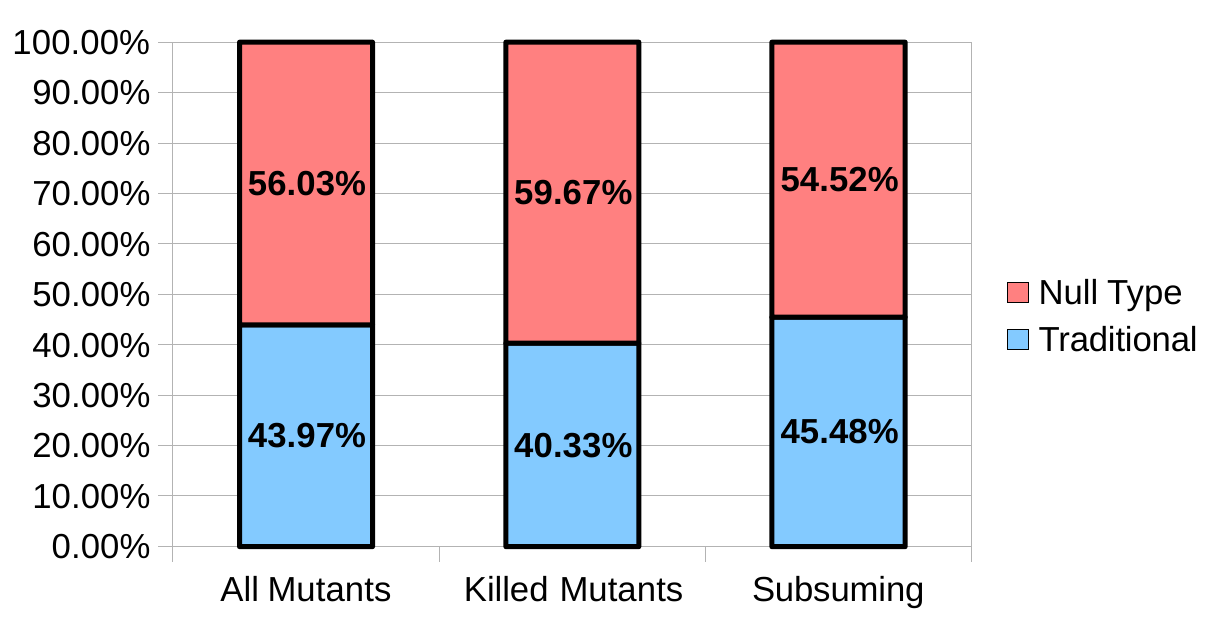}
                \caption{Ratio of Null-Type and Traditional Mutants in All, Killed, and Subsuming}
                \label{fig:mutationoperatorssubsuming-overall}
            \end{wrapfigure}

\textit{Killed mutants.}  Considering all projects as a whole, the number of generated mutants is  104,978. 
Out of this total, the number of killed and subsuming mutants are 65,551 and 16,205 respectively. This means that at least 50,029 were subsumed, and thus redundant. 
To put null-type and traditional mutants in perspective, 
Figure~\ref{fig:mutationoperatorssubsuming-overall} shows  the percentages for all,  killed, and subsuming mutants for both groups. Here, we  notice
that the percentage of the null-type mutants remains similar in these three categories.
The null-type mutants have a higher impact on the semantics of the program due to being applied at the entry and exit points of a method, the branching statements, and the declaration of an object. Therefore, the fact that they comprise a higher percentage of the killed mutants is not surprising. However, it is important to note that the distribution of null-type mutants differs only 4\% in all and killed mutants.
While 60\% of the killed mutants are null-type,  they still account for almost 55\% of subsuming mutants.  
            This indicates that \textbf{the inclusion of the null-type mutants increases the mutant redundancy only marginally}.

             \begin{figure}
                 \centering
                 \includegraphics[width=0.90\linewidth]{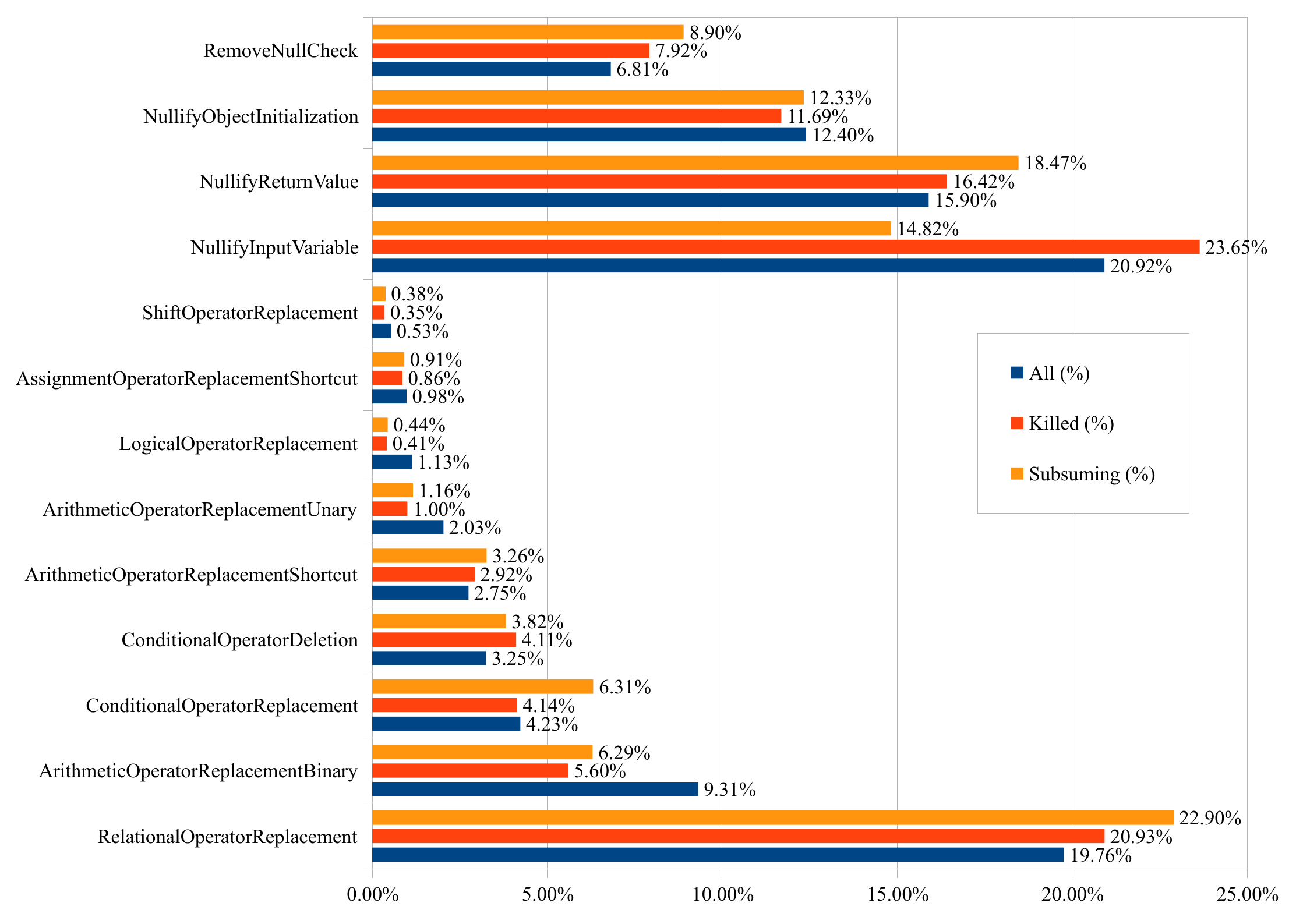}
                 \caption{Ratio of Mutants by Each Mutation Operator in All, Killed, and Subsuming}
                 \label{fig:mutationoperatorssubsuming}
                 \vspace{-1em}
                \end{figure}

To delve deeper, %
Figure~\ref{fig:mutationoperatorssubsuming} shows %
for each mutation operator the percentage of %
killed and subsuming mutants.             Among the traditional mutation operators, \textit{RelationalOperatorReplacement} and \textit{ConditionalOperatorReplacement} produce the most subsuming mutants.
The rest of the mutation operators create mutants that have the same distribution among subsuming and killed mutants. 
As this figure shows, the marginal increase in redundancy by the null-type mutation operators can be blamed on \textit{NullifyInputVariable} mutation operator. This mutation  operator produces mutants that are easier to kill compared to other mutation operators %
(21\% of all, 24\% of the killed), and  %
more of  these mutants are redundant compared to others (24\% of killed, only 15\% of subsuming). On the contrary, \textit{NullifyReturnValue} is producing fewer redundant mutants, which confirms our previous observation.

Given the results of RQ2, we can conclude that \textbf{while the inclusion of the null-type mutation operators increases the redundancy marginally, they complement the traditional mutation operators in their role of strengthening the test suite against null-type faults.}

     	\vspace*{-1em}

\section{Threats to Validity}
\label{S:Threats}
 	\vspace*{-1em}

To describe the threats to validity we refer to the guidelines reported by Yin~\cite{Yin2003}. 
Threats to \textbf{internal validity} focus on confounding factors that can influence the obtained results. 
These threats stem from potential faults hidden inside our analysis tools. 
While theoretically possible, we consider this chance limited. The tools used in this experiment have  been  used previously in several other studies, and their results went through many iterations of manual validation.
In addition, the code of LittleDarwin and LittleDarwin-Null along with all the raw data of the study is publicly available for download in the replication  package~\cite{RepPack}.

Threats to \textbf{external validity} refer to the generalizability of the results. 
In RQ1 we advocate for the adoption of null-type mutation operators by using a didactic project.  
We alleviate the  non-representativeness of this project, by analyzing  15 real open-source projects in RQ2. 
Although our results are based on projects with various levels of test adequacy in terms of traditional and null-type mutation coverage, we cannot assume that this sample is representative of all Java projects.
We use 
PITest, LittleDarwin, and Javalanche as mutation testing tools. We cannot assume that these tools are 
representative of all mutation tools available in literature. 
For this reason, we refer to these tools as general-purpose since they can work with little effort on many open-source projects.
We modeled null-types mutation operators upon the typical null-type faults described by Osman et al.~\cite{Osman2016}. 
However, there may be other types of null-type faults that we did not consider. 
Even if this was the case, our results should still hold since we already demonstrate with four mutation operators
 that they are in need of explicit modeling.

Threats to \textbf{construct validity} are concerned with how accurately the observations describe the phenomena of interest. The problem of equivalent mutants affects the analysis of surviving mutants on the test suites of the 15 open-source projects. Due to the large number of created mutants, it is impractical to filter equivalent mutants in the final results. %
Still,  we believe this threat is minimal, because we analyze two different aspects of mutation testing, which lead to converging results. %
The total number of generated mutants can be different based on the set of mutation operators that are used in each tool. However, this difference has been taken into account when discussing the results of the experiments.
To measure redundancy among the mutants, we use dynamic subsumption relationship. However, the accuracy of the dynamic subsumption relationship depends on the test suite itself. This is a compromise, as the only way to increase the accuracy is to have several tests that kill each mutant, which is not practical.

 	\vspace*{-1em}

\section{Conclusion}
\label{S:Conclusion}
 	\vspace*{-1em}

Developers are prone to introduce null-type faults in Java programs. Yet, there is no specific approach devoted to helping 
developers  strengthen the test suite against these faults. On the one hand, mutation testing provides a systematic method to create tests able to prevent common faults. On the other hand, the general-purpose mutation testing tools available today do not model null-type faults explicitly by default.

In this paper, we advocate for the introduction of null-type mutation operators for preventing null-type faults.
As a first step, we show that traditional mutation operators are not enough to cope with null-type faults as they cannot lead to the creation of a mutation-adequate test suite that can kill all of them. Then we demonstrate, by means of code examples, how the null-type mutants can drive the extension of the test suite. Finally, we highlight that null-type mutation operators are helpful in practice by showing on 15 open-source projects that real test suites %
are inadequate in detecting null-type faults. In this context, we explore the trade-offs of having null-type mutants.
On the downside, we show that the inclusion of null-type mutants increases the mutant redundancy. 
Yet, this increment is only marginal.
On the upside, we show that null-type mutants %
complement traditional mutants in two ways. 
First, they provide a large number of survived mutants to the developer to strengthen the test suite.
Second, they comprise a large part of subsuming mutants.

As a consequence, developers can increase their confidence in the test suite regarding to the null-type faults by (i) prioritizing the classes that have a large difference in traditional and null-type mutation coverage,
(ii) creating tests to kill the survived null-type mutants in these classes, and 
(iii) repeating the process until all classes have similar levels of traditional and null-type mutation coverage.

\bibliographystyle{splncs04}
\bibliography{cited}

\end{document}